\documentclass[aps, pre, twocolumn, amsmath, superscriptaddress,showkeys,showpacs]{revtex4-1}
\usepackage{graphicx}
\usepackage{color, soul}
\usepackage{dcolumn}
\usepackage{bm}
\usepackage{epstopdf}
\usepackage{caption}
\usepackage{subcaption}

\usepackage{ragged2e}
\usepackage{hyperref}
\DeclareCaptionJustification{justified}{\justifying}
\begin{document}	
	\title{Extreme events in a network of heterogeneous Josephson junctions}
	\author{Arnob Ray}
		\affiliation{Physics and Applied Mathematics Unit, Indian Statistical Institute, 203 B. T. Road, Kolkata 700108, India}
	\author{Arindam Mishra}
	\affiliation{Physics and Applied Mathematics Unit, Indian Statistical Institute, 203 B. T. Road, Kolkata 700108, India}
		\affiliation{Centre for Mathematical Biology and Ecology, Department of Mathematics, Jadavpur University, Kolkata 700032, India}
	 \author{Dibakar Ghosh}
	 		\affiliation{Physics and Applied Mathematics Unit, Indian Statistical Institute, 203 B. T. Road, Kolkata 700108, India}
	\author{Tomasz Kapitaniak}
	\affiliation{Division of Dynamics, Faculty of Mechanical Engineering, Lodz University of Technology, 90-924 Lodz, Poland}
	\author{Syamal K. Dana}	
		\affiliation{Centre for Mathematical Biology and Ecology, Department of Mathematics, Jadavpur University, Kolkata 700032, India}
		\affiliation{Division of Dynamics, Faculty of Mechanical Engineering, Lodz University of Technology, 90-924 Lodz, Poland}
	\author{Chittaranjan Hens}
		\affiliation{Physics and Applied Mathematics Unit, Indian Statistical Institute, 203 B. T. Road, Kolkata 700108, India}
	\date{\today}
	\begin{abstract}
	We report rare and recurrent large spiking events in a  heterogeneous network of  superconducting Josephson junctions (JJ) connected through a resistive load and driven by a radio-frequency (rf)  current in addition to a constant bias.  The intermittent large spiking events  show characteristic features of extreme events (EE) since they   are larger than a  statistically defined significant height.  Under the influence of repulsive interactions and an impact of heterogeneity of damping parameters, the network splits into three sub-groups of junctions, one in incoherent rotational, another in coherent librational motion and a third sub-group originating EE.  We are able to scan the whole population of junctions with their distinctive individual dynamical features either in EE  mode or non-EE mode in parameter space.  EE  migrates spatially from one to another sub-group of junctions depending upon the  repulsive strength and the damping parameter. For a weak repulsive coupling, all the junctions originate frequent large spiking events, in rotational motion when the average inter-spike-interval (ISI) is small, but it  increases exponentially with repulsive interaction; it largely deviates from its exponential  growth at a break point where EE triggers in a sub-group of junctions.  
  The probability density of inter-event-intervals (IEI) in the subgroup  exhibits a Poisson distribution.  EE originates via  bubbling instability of in-phase synchronization. 
    	\end{abstract}
	\maketitle
A typical oscillation in a pendulum  can be classified into two disparate dynamical characteristics: a rotational motion when the  trajectory of the pendulum traces the entire periphery of a  cylindrical phase space, and  a librational motion  when it is localized and never  encircles the cylindrical space \cite{Strogatz, Baker}. The pendulum motion has a superconducting analog in the classical resistively-capacitively-shunted Josephson junction (RCSJ) \cite{Levi, Duzer}. 
The RCSJ reveals spiking and bursting oscillations (all rotational motion) \cite{Dana1, Dana2, Hens} for a wide range of damping parameters and above a critical value of external DC bias  current. A  network of JJ connected to a load resistance unveils a rich diversity of collective behaviors depending upon the network structure and the coupling function.  Synchronization was noticed \cite{Wiesenfeld} long back in an array of disordered or heterogeneous Josephson junctions under a  mean-field interaction via the voltage variable.  A traveling wave pattern of bursts was reported \cite{Ustinov} in an experiment with a regular ring of JJs locally connected through the phase variable. A network of identical junctions interacting repulsively via the voltage variable was shown \cite{Mishra} to undergo a transition from coherent librational to coherent rotational motion with intermediate cluster states and chimera states for increasing repulsive interaction strength.  An induced heterogeneity in this network of junctions triggers EE in a sub-population of junctions, which is our interest of this study.
\par EE appear as significantly large amplitude events in a state variable  or in an observable of  dynamical systems, in general, and  it emerges in  phase space of the system when the trajectory of the system has been occasionally kicked out, from its usual bounded attractor phase space,  by a region of instability  lying somewhere in state space of the system. These events are sparse and occurring not too often in time domain, but recurrent. Naturally, the distribution of  events distinctly deviates from the traditional Gaussian distribution 
revealing  unusual (rare) long-tail behaviour. This type of phenomenon is common  in many natural and man-made systems \cite{Ghil, holgar}, but devastating for life and society. From a dynamical system point of view, it reveals a sudden  and sharp expansion of an attractor at a critical parameter value due to an interior crisis \cite{single, Rajat, Kingston, masolar} of the considered system. EE is seen in multistable systems where noise induces occasional attractor-hopping between between  coexisting states \cite{noise}. In a semiconductor laser experiment \cite{masolar_2}, an optical feedback  was seen to trigger  significantly high amplitude optical pulses  that demonstrate characteristic features of EE. 
On the other hand, in coupled systems, a local instability of synchronization originates EE, which is reflected  in the synchronization error of complete synchronization as  bubbling transitions \cite{prl, cs,cs1}.  EE have  also been observed due to instability of   in{-}phase \cite{ps} and  anti{-}phase synchronization \cite{aps}  of  coupled systems. 
A ubiquitous  presence of EE drives researchers to focus on  possibilities of prediction and control of such events in experiments \cite{masolar, prl} as well as  theoretical  studies of complex  systems  \cite{prediction,control}. 
\begin{figure}[h!]
	\includegraphics[scale=0.45]{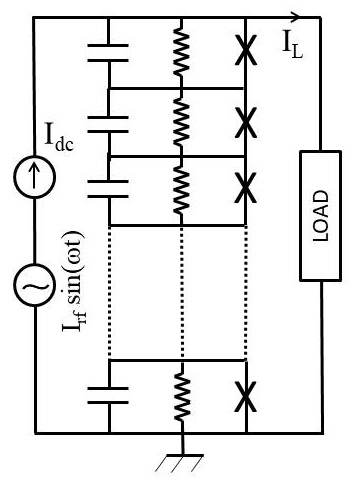} 
	\caption{Schematic diagram of a network of RCSJs in an array interacting via a common resistive load. A common DC bias current $I_{dc}$ and a rf-signal $I_{rf}sin(\omega t)$ forcing.} 
	\label{fig1} 
\end{figure}
\begin{figure*}[ht]
	\centerline{
		\includegraphics[scale=0.45]{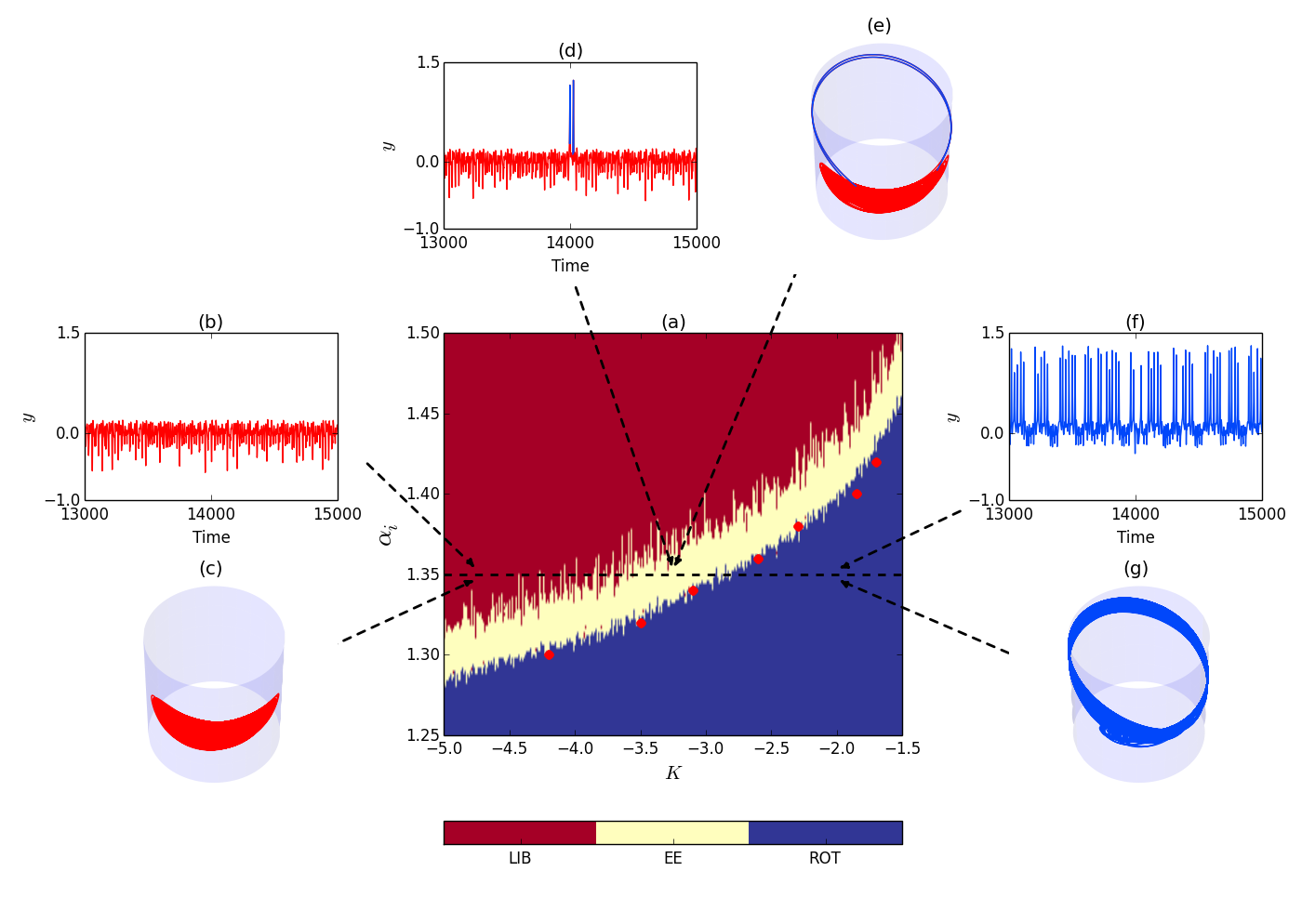} 	}
	\caption{{\bf Location of sub-groups of junctions in rotational (blue region), librational motion (red region) and EE modes (yellow band) for varying $K$ in the network of JJs.} Junction nodes are arranged (a) along the ordinate in descending order top to bottom from high to low  $\alpha_i$  values. Time evolution of the voltage variable $y$  (b,d,f) and  phase portraits in cylindrical plane (c,e,g) are presented for three junctions from different regions indicated by dashed arrows. Closed circles (red circles) denote critical values of $\langle \rm{ISI} \rangle$ that marks the onset of EE. They fall almost closely on the borderline of junctions in rotational and EE mode.} 
		\label{fig2} 
\end{figure*}
\par  EE was observed in networks of systems \cite{Rajat} as a collective response of the ensemble; no information was available on the individual dynamics of the nodes. In this rapid communication, we report emergence of EE in a network of superconducting heterogeneous RCSJs connected  in an array  in presence of  a repulsive  mean-field interaction. The key question we raise here is which population (whole or a sub-population) of the superconducting junctions  will respond to induced heterogeneity in damping and be affected by the repulsive interaction to originate EE. What is the range of coupling parameter? Most importantly, what is the mechanism behind such events (generation and termination of  EE) in the backdrop of a transition from collective rotational to librational motion with repulsive coupling? We address the questions here with affirmative answers. 
\par The network of junctions, by an interplay of distributed damping parameters and repulsive mean-field interactions, instead of attaining an expected homogeneous state \cite{Wiesenfeld}, splits into three groups: 
one sub-group of junctions is unaffected by the coupling and remains in an incoherent rotational motion, a second sub-group  transforms into librational motion in an  in-phase synchronization state and a third sub-group of junctions maintains in-phase synchrony in librational motion with the second sub-group, most of the time, but  occasionally loses stability of  synchronization. This occasional loss of synchrony is reflected in the temporal evolution  of the voltage variable of the third sub-group of junctions as rare large spiking events; the large events are classified as EE when they are larger than a statistically defined significant height or a threshold level. The interval between two threshold-crossing successive large events is defined as IEI. Most interestingly, EE occurs in an incoherent manner in the sub-group of junctions and migrate from one to another sub-population of junctions, basically from  high to low damping nodes when  the strength of  repulsive mean-field interaction is increased (see the SI \cite{suply}). This is a generic behavior of the heterogeneous RCSJ array under repulsive  mean-field interaction irrespective of the network size and  coupling strength.  
\par The network of  RCSJs in an array connected via a common resistive load \cite{Wiesenfeld} is shown in Fig.\ref{fig1}.  The circuit equations of the network can be easily written  using the Kirchoff's laws. The dynamics of the $i$-th node of the rf-forced junction array is then represented by non-dimensionalized  evolution equations of phase $\phi_i$ and voltage $y_i$,
	\begin{eqnarray}
		\dot{\phi_{i}} & =& y_{i}\\
		\dot{y_{i}} &=& I_{dc} - \sin \phi_{i} - \alpha_{i} y_{i} + I_{rf}\sin (\Omega_{rf}t) + KY.
	\end{eqnarray}
	where  $Y$= $\frac{1}{N} \sum\limits_{j=1}^N  {y_j}$, 
	$\alpha$=$[h/2 \pi e I R^2 C]^{1/2}$$ 
	= (\frac{1}{\beta})^{\frac{1}{2}}$ 
	is the damping parameter, 
	$\beta$ 
	is the McCumber parameter, 
	$I_{dc}$ 
	is the normalized constant bias current. Here, $R$ and $C$ are intrinsic resistance and capacitance of each junction, respectively. 
	$K$ defines the strength of mean-field interaction between the junctions. 
We search for an appropriate  parameter space in a network of $200$ junctions for the emergence of EE. The  constant bias for each RCSJ node is fixed at  $I_{dc} = 1.2$ when an uncoupled junction is in rotational motion \cite{Hens}. 
The amplitude and frequency of the radio-frequency ($rf$) signal is set at 
$I_{rf} = 0.26$ and
$\Omega_{rf} = 0.4$.  We have checked that the phenomenon is not restricted to this choice of parameters only. The damping parameters 
$(\alpha_{i}, i=1,2,...,N)$ 
are uniformly distributed in a range of $1.25$ to $1.50$ values, i.e., when all the junctions show  spiking rotational motion, in isolation.  The impact of heterogeneity of damping on  the emergence of EE in the network of junctions has been numerically explored here for a repulsive interaction. 
The repulsive mean-field interaction occurs naturally in the junction network due to an outgoing current flow across the load (following Kirchoff's law) and it is a necessary condition for the origin of EE. 
\par Our main results are demonstrated in  Fig.\ \ref{fig2}(a) where the ordinate  represents the  junction index ($i$) arranged  in an ascending order of damping parameter ($\alpha_i$)  so that a junction with a lowest damping ($\alpha_i$) is placed at the bottom of the axis. 
A yellow (white) band forms in the middle region 
that describes a sub-group of junctions in EE mode. This  band separates two other regions representing two subgroups of junctions, in non-EE mode, one  in coherent librational motion (red region) and the other in incoherent rotational motion (blue region) in Fig. \ref{fig2}(a). 
For elaboration, we arbitrarily  select a junction with a  damping 
$\alpha=1.35$ 
 from the whole population and investigate the nature of trajectories for three cases (horizontal black dashed line) by  changing $K$ from large to smaller repulsive strength along the abscissa,  (i) for $K=-4.5$ 
 (red region), the junction node is in librational mode as shown in the time evolution in Fig.\ \ref{fig2}(b) and its attractor in a cylindrical phase space in Fig.~\ref{fig2}(c). This is an emergent property of the junction and counterintuitive to its  isolated  dynamics which was originally rotational now suppressed by the repulsive interaction. For a choice of intermediate repulsive strength (ii) 
 $K=-3.5$ (yellow band), 
 the junction  enter the EE mode; an exemplary  big spiking event (blue) is shown in a short time run in Fig.\ \ref{fig2}(d) when its  trajectory remains confined, most of the time, to localized librational motion (dense red region), but occasionally visits  the periphery of the cylindrical phase space (blue line) as shown in Fig.\ \ref{fig2}(e), which  shows a signature of EE in a long run.  For 
 $K=-1.8$ (blue region), 
this junction shows up highly irregular, but very frequent spiking and bursting in rotational motion in Figs.\ \ref{fig2}(f)-(g) that fails to qualify as EE. Alternatively, if we look at the junctions' dynamics at a constant $K$ value (along the vertical axis), it reconfirms  that EE appear in a subgroup of junctions (yellow band), but this subgroup changes  location with $K$. In a sense, the junction population infected by EE migrate with $K$ (see the supplementary section \cite{suply} for more details). This is a generic behavior of a heterogeneous RCSJ array under repulsive  mean-field interaction irrespective of the network size and  coupling strength.   The borderline between the subgroups of junctions in rotational (blue color) and the EE regions (yellow color) is traced by using a measure of significant height ($H_s$); a junction, in rotational motion, crosses the border to be included in the EE mode for larger repulsive strength when the spiking events become rarer and larger than the  $H_s$ level.  This borderline is  further demarcated by measuring a critical value of the average inter-spike-interval $\langle \rm{ISI} \rangle$ of large frequent spikes in seven arbitrarily selected junctions, in rotational motion.  The critical values of $\langle \rm{ISI} \rangle$   the seven junctions are marked by closed circles (red color) as shown in Fig.\ \ref{fig2}(a), which fall closely on the borderline. The critical 
$\langle \rm{ISI} \rangle$  value
 is explained in detail in Fig.\ \ref{fig3}(e). 
\begin{figure}[ht]
	\centerline{
		\includegraphics[scale=0.45]{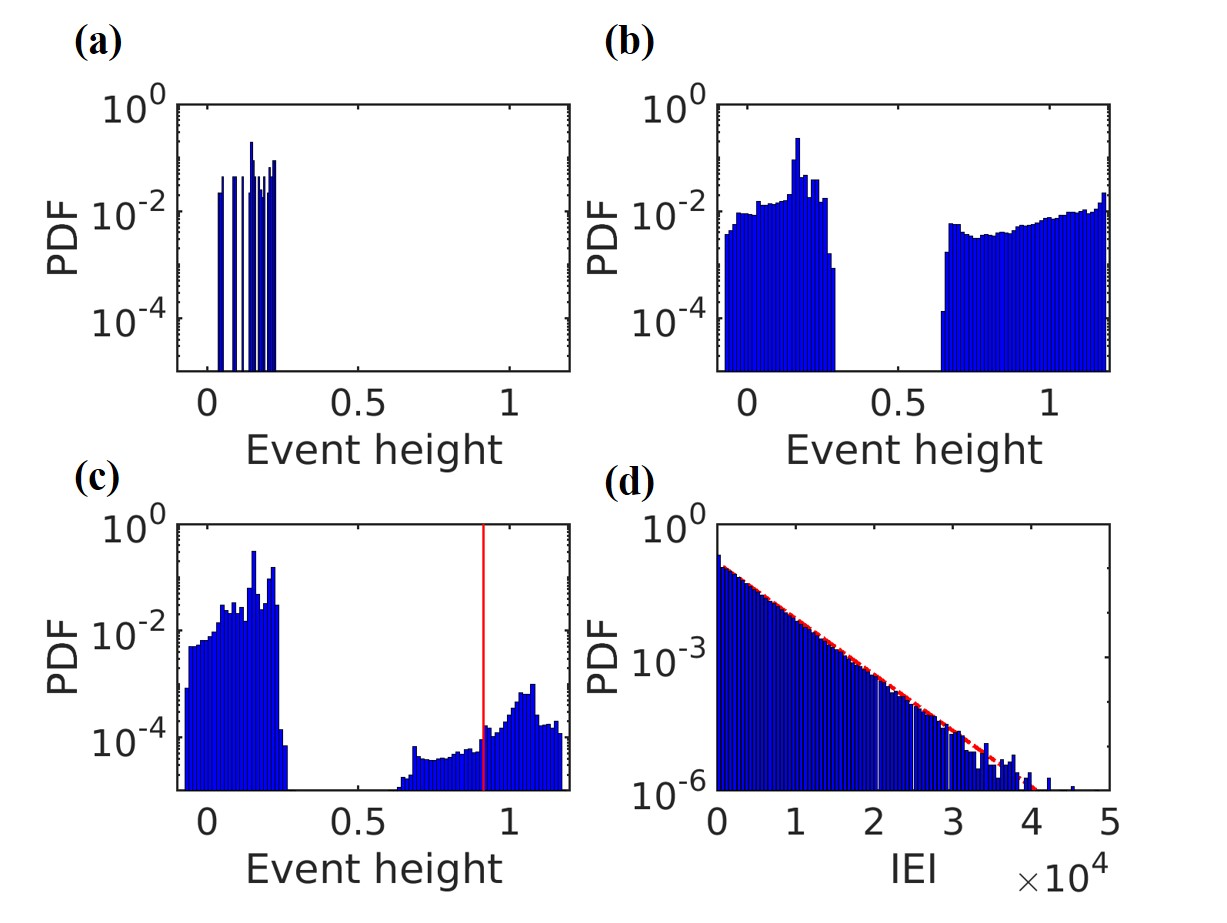} }
	\caption{{\bf Distribution of event heights and IEI.}  (a) PDF for librational motion: a localized distribution ($K=-4.5$), (b) PDF for rotational motion: distributed around two centres, essentially distributed  in the whole cylindrical space ($K=-1.8$), (c)  PDF of  EE:  crosses the threshold $Hs$ (red vertical line), (d) PDF  of IEI indicates exponential distribution ($K=-3$) with a linear fit (red dashed line).}
	\label{pfig2} 
\end{figure}
\par 
For a quantitative classification of EE, we consider the local maxima or peaks as events in the time evolution of voltage  variable ($y$) to define the significant height as an EE qualifier,  
$H_s$=$\mu$+$8\sigma$, 
where $\mu$  is the average height of all the peaks in a measured time series (data recorded from a long numerical run) and 
$\sigma$ is the standard deviation of the peaks \cite{muller, ocean, masolar, masolar_2, Kingston}. 
For a further distinction of EE from nominal  events, we estimate  the probability density function (PDF) of event heights.  Firstly, we estimate PDF of events in a  junction selected from the subgroup in complete libration where the distribution is confined to small size events with a limited spreading of height except around zero as shown in Fig.\  \ref{pfig2}(a). On the other hand, PDF of event-heights in junctions in rotational motion is shown in Fig.\  \ref{pfig2}(b), 
which is almost equally distributed between librational (left hand side) and rotational motion (right hand side), except a gap around $0.5$. No event of intermediate size really appears [cf., Fig.\ \ref{fig1}(f)] and this shows a gap in the middle.
In this state, the switching from a small amplitude librational state to rotational motion is so frequent that the spiking events has a very large $H_s$ value and no spiking event qualifies as an EE.  
PDF of event heights in a junction in EE mode (chosen from the yellow band) in Fig.\ \ref{fig1}(a), shows a dramatic change in the distribution in Fig.\ \ref{pfig2}(c); event-heights around $1$
are much less probable (compared to event-heights around $0$). Transition to rotational motion is now occasional, confirming emergence  of rare and recurrent events, which exceed a pre-assigned threshold 
$Hs=0.9143$ (vertical red line) estimated from event-heights of our simulated data. 
We also estimate IEI of successive EE in a long run and plot their PDF in Fig.\ \ref{pfig2}(d) that follows  a Poisson distribution \cite{Ghil}   re-confirming the origin of rare  uncorrelated events (cross-checked separately, but not presented here) in the network of junctions.
PDFs is fitted by 
$P(r) = \lambda exp(- \lambda r)$,
where $r$ is the IEI and $\lambda$ is the shape parameter whose estimated value is 
$\lambda= 0.0003$.
\begin{figure*}[ht]
	\centerline{
		\includegraphics[scale=0.7]{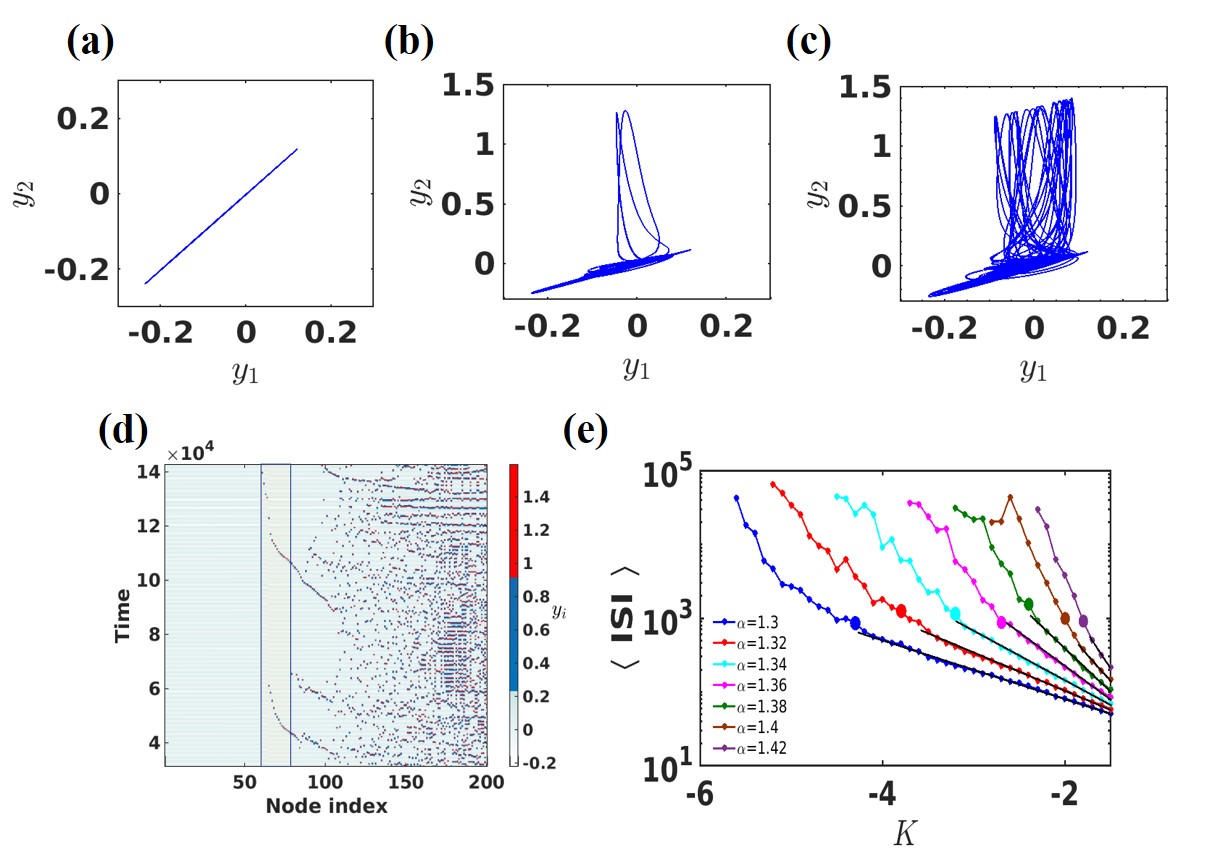} 	}
	\caption{{\bf  Synchronization, spatio-temporal pattern of junctions and average  inter-spike-interval.} $y_1$ vs. $y_2$ $(i=1,2)$ plots of two junctions (a) arbitrarily selected both in librational motion that maintain in-phase synchrony, (b) one in  librational motion and another in EE; two bubbles indicate instability of the synchronization manifold along  the transverse direction, signifying two occasional large events, (c) one node in rotational, another in  librational motion; many bubbles indicating strong instability in the synchronization manifold.  (d) Spatio-temporal dynamics of the entire network for a constant $K=-3$;  
	light blue region indicates coherent librational mode in $59$ junctions, 
	an even smaller sub-population between two vertical black lines, shows signatures of EE of varying heights (red and blue dots), a larger population at right of the vertical lines are in rotational motion; large spikes (blue and red dots) are very frequent, (e) $\langle \rm{ISI} \rangle$ estimated for seven arbitrarily selected  junctions for a range of coupling $K$.}
					\label{fig3} 
\end{figure*}
\par  The triggering of EE in any junction is linked here to the origin of instability in the  synchronization manifold of the connected junctions. 
For illustration, we present examples of two junctions selected from three different regions of Fig. \ref{fig1}(a) for  a fixed $K=-3$. 
When two nodes are chosen from the libration region (red color), they are seen to follow strong in-phase synchronization (almost complete synchronization) as shown in Fig.\ \ref{fig3}(a). Figure  \ref{fig3}(b) shows a comparative scenario of two junctions, one chosen from libration regime (red color) and another from the EE population (yellow color). It confirms that the junction nodes  maintains in-phase  synchrony for most of the time (deep blue line),
however, instabilities are seen along the transverse direction to the synchronization manifold and shows signatures of EE. The deflection along the transverse  direction (two bubbles appear vertically from in-phase manifold) appear as EE as shown in Fig.\ \ref{fig3}(b). 
Any junction node  in rotational motion (blue region) with any other node in a librational or EE node is in a desynchronized state (many bubbles)  as shown in Fig.\ \ref{fig3}(c). To elucidate the entire observation  with more details, we make a spatio-temporal plot in Fig.\ \ref{fig3}(d).  First $59$ 
nodes (left of abscissa) are in librational motion (white and light  blue color)  and clearly follow in-phase synchronization; a coherent pattern is seen in spatial as well as temporal domain. Last 
 $122$ 
nodes on the right side are in rotational motion (red and blue color together confirm rotational motion) and they do not follow any specific pattern in the spatio-temporal domain; they are desynchronized to each other. A small population of  intermediate junction nodes (demarcated within  two vertical black lines) are in EE mode when large spikes are rare (red and blue dots are sparse).  Clearly, EE never emerge coherently in the junctions (see blue and red dots within the vertical lines) in the sub-group. However, the junctions, in this domain, follow in-phase correlation with librational nodes except at the time of emergence of large spikes.  Another noticeable feature in the spatio-temporal domain is the number of large spikes (red and blue dots  in the backdrop of white and light blue color) that  increases  continuously at the right side.  This is  further confirmed by plotting  $\langle \rm{ISI} \rangle$   as a function of repulsive coupling strength (semi-log plot) in Fig.\ \ref{fig3}(e) for seven selected nodes in rotational motion. One particular node  with lower 
$\alpha=1.3$ 
selected to be frequently spiking in rotational motion, show linear increase in $\langle \rm{ISI} \rangle$ (small blue circles) with $-K$ indicating an exponential growth until it reaches a critical value near a break point (big blue circle) when it makes a sudden change in the behavior deviating largely from a linear fit (black line). At this break point, the dynamics is no  more rotational, however, rare transition to this mode occurs indicating an onset of EE.  A closed circle (blue color) denotes the onset point of EE in the junction when $K$ value corresponds to the 
border of yellow and blue regions in Fig.\ \ref{fig1}(a).  All the other six junctions  (red, cyan, magenta, green, brown and violet color lines) follow the same trend. Clearly, a node with a lower damping ($\alpha =1.30$) needs more repulsive coupling strength (big blue circle  on the blue line)  for transforming its rotational motion into EE. As we increase the damping, a low repulsive strength is necessary for the onset of EE. For each of the seven cases,  $\langle \rm{ISI} \rangle$ exponentially  increases until it reaches a critical value near the onset of EE  at a breaking point; once EE starts, the slope of $\langle \rm{ISI} \rangle$ is extremely enhanced leading to rare occurrences of spikes.  The
linear fit of $\langle \rm{ISI} \rangle$, 
in the weaker repulsive range of coupling,  can work as a  predictor of EE  since linearity deviates significantly at the break point. 
\par We studied a network of heterogeneous Josephson junctions under repulsive mean-field interaction that showed extreme-events-like dynamical features in a subpopulation of the junctions. 
Starting with a network of junctions with distributed damping parameters that originate frequent large spiking events in rotational motion, we studied the impact of a natural repulsive mean-field interaction by applying a common resistive load. The junctions in the network, instead of attaining an expected  homogeneous state, splitted into three dynamical subgroups for a range of repulsive strength: one  subgroup of junctions remained in rotational motion and desynchronized, a second subgroup of junctions switched over to librational motion in a state of in-phase synchrony and a third subgroup of smaller number of junctions showed rare transition from librational to rotational motion revealing signatures of EE. This third sub-group of junctions showed in-phase synchrony with the second subgroup in librational motion when the junctions were restricted to small amplitude oscillation, but occasionally moved away to bubble out of the in-phase synchronization  manifold and originated large spiking events henceforth characterized as EE. PDF of all such events showed rare occurrence of EE and the IEI followed a Poisson distribution. $\langle \rm{ISI} \rangle$ was estimated for seven selected junctions in rotational motion for weak repulsive coupling when it was increasing exponentially with repulsive mean-field strength.  
It deviated largely  from its exponential growth at a break point indicating a critical value of repulsive strength when we saw onset of EE in each of the seven junctions. We were able to locate the sub-population of junctions where EE may emerge with varying  repulsive strength of interaction. EE never remained confined  to a fixed subpopulation, but migrates from a group of high to low damping nodes in the network for an increase of repulsive strength \cite{suply}.  
\par T.K. has been supported by the National Science Centre, Poland, OPUS Programme Project No. 2018/29/B/STB/00457. C.H. is supported by INSPIRE-Faculty grant (code: IFA17-PH193).

\end{document}


\title{Extreme events in a network of Josephson junctions}
{\tiny {\author{Arnob Ray$^1$, Arindam Mishra$^{1,2}$, Dibakar Ghosh$^1$, \\Tomasz Kapitaniak $^{3}$, Syamal K Dana $^{2,3}$,\\  Chittaranjan Hens $^1$\\
$1.$ Physics and Applied Mathematics Unit, Indian Statistical Institute, \\203 B. T. Road, Kolkata 700108, India\\
$2.$ Centre for Mathematical Biology and Ecology, Department of Mathematics,\\ Jadavpur University, Kolkata 700032, India\\
$3.$ Division of Dynamics, Faculty of Mechanical Engineering, \\Lodz University of Technology, 90-924 Lodz, Poland}}	}
\date{}
\maketitle
\begin{figure*}[ht]
	\centerline{
		\includegraphics[scale=0.75]{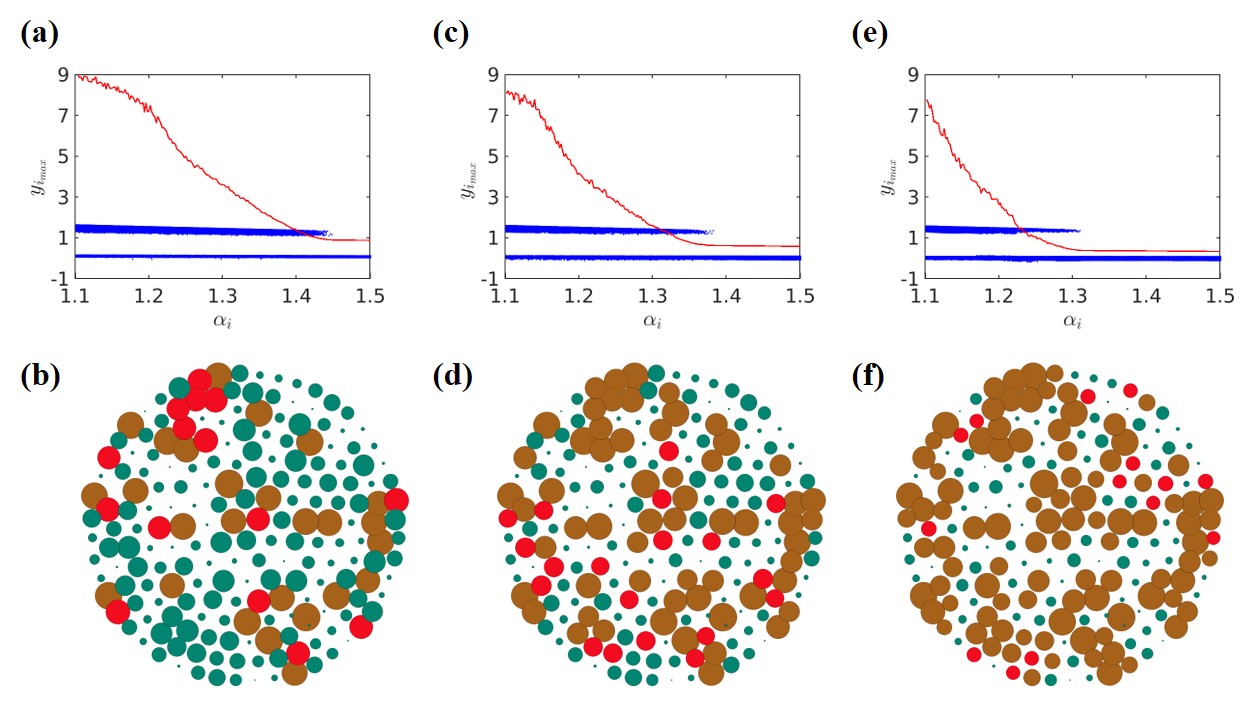}}
	\caption{{\bf Emergence of EE and  migration of EE inside the network.} Subfigures (a), (c) and (e) trace the bifurcation like diagram as a function of node index in the network for three negative strengths: $K=-1.8, -3.0,$ and $-5.0$, respectively.  In (b), (d) and (f), the brown nodes are in librational motion, the deep green nodes are in rotational motion, whereas the red nodes reveal rare events, i.e., EE. The size of the nodes is specified by their damping value, if the damping is high, the size of the node is also high. With further increase of the negative strength, the nodes having rare events (shown by red color in (b), (d) and (f)) move  from higher to lower damping.}
	\label{fig4} 
\end{figure*}

Now, we look at the pattern of appearance of extreme events in the complete graph of non-identical Josephson Junctions from a different viewpoint. If we fix the repulsive coupling at a lower strength, an interesting feature occurs within the network. The higher damping nodes quickly absorb the weak repulsive mean-field  and  the trajectories in the periphery of the cylinder are pushed inward to remain quiet in a  localized space  creating librational motion. However, occasionally few trajectories of some of the higher damping nodes come to the periphery of the cylindrical phase space to trace a rotational motion by an interplay of high damping and repulsive interaction. An occasional appearance of trajectory in the periphery originates the EE of those nodes. Meanwhile, the lower negative strength is too weak to destroy the rotational motion of   the lower damping nodes. Therefore, we conclude that a mixed population in the repulsive environment can reveal rare events in some of the nodes of the network. These features are shown in the Figs.\ \ref{fig4}(a) and \ref{fig4}(b). The  $x$-axis shows the ascending order of the damping parameter $\alpha$ of all the nodes. In $y$-axis, the local maximum of each node is plotted. For weak coupling strength at $K=-1.8$, the higher damping  oscillators  ({\it i.e.}, $\alpha_i>\sim1.42$, nodes with bigger size in Fig.\ \ref{fig4} (b)) 
are pushed towards librational motion (describe by brown filled circle), whereas the lower damping oscillators
({\it i.e.}, $\alpha_i<\sim1.40$, nodes with smaller size in Fig.\ \ref{fig4} (b)) are still in rotational motion (green filled circle). However, few intermediate nodes (medium size is shown with red filled circle) pass the pre-defined threshold $T$ (page 3 in main text), 
therefore reveal EE. If we increase the negative strength 
($K=-3.0$), the more oscillators with higher damping undergo librational motion, and the number of nodes having rotational motion decrease (Figs.\ \ref{fig4}(c) and \ref{fig4}(d)), however the EE still emerge in few of the intermediate nodes in the network.
By further increasing the negative strength
($K=-5.0$),
  more number of nodes  are pushed towards the localized space by losing their rotational motion. Therefore, the rare events are revealed in near to the lower damping nodes (Figs.\ \ref{fig4}(e) and \ref{fig4}(f)).  We can say here the extreme event is migrating from one  to the other set of nodes (from higher damping to lower damping,  red nodes) if we increase the mean filed  strength from lower negative to higher negative.  The phenomena can be explained as
 follows: the node having weak damping resists the negative strength and maintains its rotational motion. The
 more negative strength is required to transform its rotational motion to librational motion as the trajectories
 of this node can be completely perturbed, therefore it's  rotational motion is suppressed to localized librational
 motion. In intermediate strength, the trajectories of that  oscillator occasionally visit the entire periphery in the
 cylindrical space and EE occurs on that range of coupling  (follow the horizontal dashed black line in Fig. 1(a) in the main text).    
  Note that  the EE phenomena occur  only if there is a transition between rotational motion to the librational motion or vice-versa. 
